\documentclass[]{emulateapj}
\usepackage{color}

\renewcommand\thefootnote{\alpha{footnote}}

\newcommand{\Hp}{\hspace{1mm}}
\newcommand{\Suzaku}{{\it Suzaku}}

\shorttitle{{\it SUZAKU} OBSERVATIONS OF NGC~1566 AND NGC~4941}
\shortauthors{Kawamuro et al.}

\begin{document}
\title{BROAD BAND X-RAY SPECTRA OF TWO LOW-LUMINOSITY ACTIVE GALACTIC
NUCLEI NGC~1566 AND NGC~4941 OBSERVED WITH SUZAKU}

\author{
 Taiki Kawamuro\altaffilmark{1},
 Yoshihiro Ueda\altaffilmark{1},
 Fumie Tazaki\altaffilmark{1},
 Yuichi Terashima\altaffilmark{2}
}

\altaffiltext{1}{Department of Astronomy, Kyoto University, Kyoto 606-8502, Japan}
\altaffiltext{2}{Department of Physics, Ehime University, Matsuyama 790-8577, Japan}

\begin{abstract} 

We report the first broad band X-ray spectra of the low luminosity
active galactic nuclei (LLAGN) NGC 1566 (type 1.5) and NGC 4941 (type 2)
observed with {\it Suzaku} and {\it Swift}/BAT covering the 0.5--195 keV
band. Both targets have hard X-ray luminosities of $\sim 10^{41-42}$
ergs s$^{-1}$ in the 15--55 keV band. The spectra of the nucleus are
well reproduced by a sum of partially or fully covered transmitted
emission and its reflection from the accretion disk, reprocessed
emission from the torus accompanied by a strong narrow iron-K$\alpha$
line, and a scattered component (for NGC 4941). We do not significantly
detect a broad iron-K$\alpha$ line from the inner accretion disk in both
targets, and obtain an upper limit on the corresponding solid angle of
$\Omega/2\pi < 0.3$ in NGC 1566. The reflection strength from the torus
is moderate, $\Omega/2\pi =0.45^{+0.13}_{-0.10} $ in NGC 1566 and
$\Omega/2\pi =0.64^{+0.69}_{-0.27} $ in NGC 4941. Comparison of the
equivalent width of the narrow iron-K$\alpha$ line with a model
prediction based on a simple torus geometry constrains its half-opening
angle to be $\theta_{\rm oa} \simeq 60^\circ-70^\circ$ in NGC
4941. These results agree with the obscured AGN fraction obtained
from hard X-ray and mid-infrared selected samples at similar
luminosities. Our results support the implication that the averaged
covering fraction of AGN tori is peaked at $L\sim 10^{42-43}$ ergs
s$^{-1}$ but decreases toward lower luminosities.

\end{abstract}


\section{INTRODUCTION}

X-ray surveys of active galactic nuclei (AGNs) have revealed that the
cosmological evolution of AGN luminosity function is well described by
the luminosity-dependent density evolution characterized by
``down-sizing'', where the low-luminosity AGNs show the peak of their
comoving space density at lower redshift than that of high-luminosity
ones \citep[e.g.,][]{Ued03,Has05}. It is being confirmed by optical
AGN surveys of unobscured AGNs \citep[e.g.,][]{Ike12}. This fact
suggests that the growth history of supermassive black holes (SMBHs)
may fundamentally differ between the low mass and high mass
SMBH-galaxy systems in the present-day universe. Thus, it is quite
interesting to investigate the nature of low-luminosity AGNs (LLAGNs)
in comparison with more luminous populations. 
Despite their abundance in the space density, however, the
detailed study of {\it broad band} X-ray spectra of LLAGNs has been
limited due to their faint nature \cite[for previous works below 10
keV, see e.g.,][]{Tera02,Cap06,Ho08,Emm13}.

Observed AGN properties strongly depend on the luminosities in many
aspects. In particular, it is known that the obscured (type-2)
fraction of AGNs shows strong anti-correlation with luminosity
\citep[for X-ray selected samples see e.g.,][]{Ued03,Has08},
indicating that the covering fraction of obscuring material around a
SMBH is smaller in more luminous AGNs. Recent studies based on hard
X-ray ($>15 $keV; \citealt{Bec09,Bur11}) and mid-infrared
\citep{Bri11b} selected samples suggest that this relation is more
complex when LLAGNs are included. The obscured fraction has a peak of
$\approx$0.6--0.8 around the 15--55 keV luminosity (hereafter
$L_{15-55}$) of $\sim 10^{42-43}$ ergs s$^{-1}$ below which it starts
to decrease toward lower luminosity, although the sample size of
LLAGNs is quite limited. To understand the AGN phenomena in a unified
way over the wide range of luminosity, it is crucial to establish the
trend and reveal the origin.

Broad band X-ray data are useful to constrain the geometry of
surrounding material around the SMBH as well as physics of the inner
accretion disk. The {\it Suzaku} satellite \citep{Mit07} has unique
capability to simultaneously observe X-ray spectra covering the 0.2--700
keV band with good energy resolution below 10 keV. Recent numerical
models of X-ray spectra based on Monte Carlo simulations
\citep{Ike09,Mur09,Bri11a} are powerful tools to
infer the geometry of the torus, such as the opening angle, column
density, and inclination angle \citep[e.g.,][]{Awa09,Egu11,Taz11}.

Since the obscured fraction of AGNs should reflect the covering fraction
of the absorbing material, we expect the torus structure of LLAGNs with
$L_{\rm 15-55} < 10^{42}$ ergs s$^{-1}$ may be different from that of
typical Seyfert galaxies with $L_{\rm 15-55} \sim 10^{43}$ ergs
s$^{-1}$. In this paper, we present results from the first simultaneous
X-ray spectra covering the 0.5--30 keV band of two nearby LLAGNs,
NGC~1566 ($z$ = 0.0050) and NGC~4941 ($z$ = 0.0037) observed with {\it
Suzaku}. These targets have $L_{\rm 15-55} \sim 10^{41-42}$ ergs s$^{-1}$
and were selected from the {\it Swift}/BAT AGN catalog \citep{Bau12},
which ensures sufficiently bright hard X-ray fluxes above 10 keV. Our
main aim is to constrain their torus geometry by precise measurements of
the reflection component with an iron-K$\alpha$ fluorescence line. {\it
Swift}/BAT spectra in the 14--195 keV band averaged over 70 months are
also utilized in the analysis.

%%%%%%%%%%%%%%%%%%%
% OBSERVATION LOG %
%%%%%%%%%%%%%%%%%%%
\begin{deluxetable*}{ccc}
\tabletypesize{\footnotesize}
\tablecaption{Observation Log\label{OBSERVATION_LOG}}
\tablewidth{0pt}
\tablehead{\colhead{Target Name} & \colhead{NGC~1566} & \colhead{NGC~4941}}
\startdata
Start Time (UT) & 2012 May 19 \,1:36  & 2012 June 22 11:16 \\ 
End \ Time (UT) & 2012 May 20 \,6:34 & \,2012 June 24 \,5:52 \\ 
Exposure (XIS) (ks) \tablenotemark{a}& 72.8 & 80.8 \\
Exposure (HXD/PIN) (ks) & 76.0 & 72.0 \\ 
\enddata
\tablenotetext{a}{Based on the good time interval of XIS-0.}
\end{deluxetable*}
%%%%%%%%%%%%%%%%%%%
%%%%%%%%%%%%%%%%%%%

%%%%%%%%%%%%%%%%%%%
%    INFOMATION   %
%%%%%%%%%%%%%%%%%%%
\begin{deluxetable*}{ccc}
\tabletypesize{\footnotesize}
\tablecaption{Information of Targets\label{INFO}}
\tablewidth{0pt}
\tablehead{\colhead{Target Name} & \colhead{NGC~1566} & \colhead{NGC~4941}}
\startdata
SWIFT ID & SWIFT J0420.0-5457 & SWIFT J1304.3-0532 \\
R.A. (J2000)\tablenotemark{a} & 04 20 00.42& 13 04 13.14 \\
Decl. (J2000)\tablenotemark{a}&-54 56 16.1 & -05 33 05.8\\
Redshift & 0.0050 &  0.0037\\
Classification & Seyfert 1.5& Seyfert 2 \\
\enddata
\tablenotetext{a}{The position of each source is taken from the NASA/IPAC Extragalactic Database.}
\end{deluxetable*}
%%%%%%%%%%%%%%%%%%%
%%%%%%%%%%%%%%%%%%%

NGC~1566 is classified as Seyfert~1.5 \citep{Ver06} and has a
morphology of barred spiral, Sb \citep{San94,Fis08}. No good quality
X-ray spectra have been obtained up to present even below 10 keV. NGC
1566 was detected in the {\it XMM-Newton} slew survey \citep{Sax08},
and its 2--10 keV luminosity ($L_{\rm 2-10}$) was estimated to be $L_{\rm 2-10} \sim
10^{41.5}$ ergs s$^{-1}$ for an assumed photon index of $\Gamma=1.7$
\citep{Lev09}. Using archival data of the {\it ROSAT} High Resolution
Imager, \citet{Liu05} detected four Ultraluminous X-ray sources
within NGC 1566, whose summed luminosity is $\sim 10^{40}$ ergs 
s$^{-1}$ in the 0.3--8 keV band. Thus, we neglect their
contribution in our spectral analysis. The black hole mass is
estimated to be log $M_{\rm BH}/M_\odot \approx 7.0$ through the
empirical $M$--$\sigma_\ast$ relation \citep{Gul09} with the velocity
dispersion given in the HyperLeda database \citep{Pau03}. The
Eddington ratio is calculated to be ${\rm log} L_{\rm bol}/L_{\rm Edd}
= -2.5$ from a 2--10 keV luminosity converted from the {\it Swift}/BAT
one for $\Gamma=2.0$ (our best-fit). Here we assume a bolometric
correction factor of $L_{\rm bol}/L_{\rm 2-10} = 10$
\citep{Ho09,Vas09}.

NGC 4941 is a Seyfert~2 galaxy \citep{Ver06} with a morphology type of
Sa \citep{Fis08}. In the X-ray band, this source was observed with {\it
ASCA} in 1996 July and 1997 January \citep{Tera02,Car07}, with {\it
Beppo}SAX in 1997 January \citep{Mai98}, and with {\it Chandra} in 2011
March \citep{Bot12}. The {\it ASCA} and {\it Beppo}SAX observations
reveal that the spectra are heavily absorbed with a strong
iron-K$\alpha$ emission line. As the {\it Beppo}SAX/PDS data above 10
keV are not usable due to the poor statistics \citep{Mai98}, our {\it Suzaku}
data provide the first simultaneous broad band X-ray spectra in the
0.5--30 keV band from this source. The black hole mass of NGC 4941 is estimated
to be ${\rm log} M_{\rm BH}/M_{\rm \odot} \approx 6.9$ \citep{Asm12},
and the Eddington ratio is ${\rm log} L_{\rm bol}/L_{\rm Edd} \approx
-2.4$ from the {\it Swift}/BAT luminosity with $\Gamma=1.9$ and $L_{\rm
bol}/L_{\rm 2-10} = 10$.

The organization of this paper is as follows. Section~2 describes the
observations and data reduction. The analysis and results are presented
in Section~3, and discussion is given in Section~4. We summarize our
conclusion in Section~5. Throughout the paper, we adopt distances of
16.5 Mpc for NGC~1566 and 19.7 Mpc for NGC~4941 \citep{The07} in
calculating the luminosities unless otherwise stated. In all spectral
analysis, we apply the Galactic absorption fixed at $N_{\rm H}^{\rm
gal} =8.61\times10^{19}$ cm$^{-2}$ for NGC~1566 and $
2.17\times10^{20}$ cm$^{-2}$ for NGC~4941, which are estimated from
the H~I map \citep{Kal05}. The solar abundances by \citet{An89} are
assumed in all cases. The errors attached to spectral parameters are
given at 90\% confidence limits for a single parameter of interest.

\section{OBSERVATION AND DATA REDUCTION}

\subsection{Observations}

We observed NGC 1566 and NGC 4941 with {\it Suzaku} \citep{Mit07} in
2012 May and June, respectively, for a net exposure of $\approx$80 ks
each. {\it Suzaku} is the Japan-US X-ray astronomy satellite. It
carries four X-ray CCD cameras called X-ray Imaging Spectrometer
(XIS-0, XIS-1, XIS-2 and XIS-3) as focal plane detectors of four X-Ray
Telescopes (XRTs), and a collimated hard X-ray instrument called the
Hard X-ray Detector (HXD) composed of Si PIN photodiodes and GSO
scintillation counters. XIS-0, XIS-2, and XIS-3 are
front-side-illuminated CCDs (FI-XISs), while XIS-1 is
back-side-illuminated one (BI-XIS). XIS-2 has not been available since
2007 November 7. The XISs cover the energy range of 0.3--12 keV, while
the HXD/PIN and HXD/GSO cover 15--70 keV and 50--600 keV,
respectively. Our targets were pointed close to the center of the XIS
field-of-view (the XIS nominal position). The observation log and
basic information are listed in Tables~\ref{OBSERVATION_LOG} and
\ref{INFO}.

\subsection{Data Reduction}

We analyze the {\Suzaku} data using the HEAsoft version 6.12
package, starting from the unfiltered event data produced by the
pipeline processing version 2.7.16.33. The spectral analysis is performed
on XSPEC version 12.7.1. To apply the latest energy calibration of
XIS, we reprocess the unfiltered data with {\tt xispi} and {\tt
@xisrepro} on the basis of the calibration database (CALDB) released
on 2012 September 12. 

The XIS events are extracted from a circular region with a radius of
2.8 arcmin (NGC~1566) or 1.8 arcmin (NGC~4941) around the peak of the
point spread function of the XRT. The background is taken from
source-free, circular regions within the field-of-view. We only use
the PIN data from the HXD, as our targets are too faint in the energy
band above 50 keV to be detected with the GSO. We utilize so-called
the ``tuned'' non X-ray background (NXB) model of HXD/PIN produced
with the {\tt LCFITDT} method \citep{Fuka09}. Then, the modelled
Cosmic X-ray Background (CXB) spectrum based on the formula of
\cite{Gru99} is added on that of the NXB. The systematic error of the
NXB is estimated to be $\simeq 0.34\%$ in the 15--40 keV band for a
unit of 40 ks exposure, which does not significantly affect our results.

\subsection{Light Curves}

\begin{figure}
\begin{center}
\epsscale{0.85}
\rotatebox{-90}{
\plotone{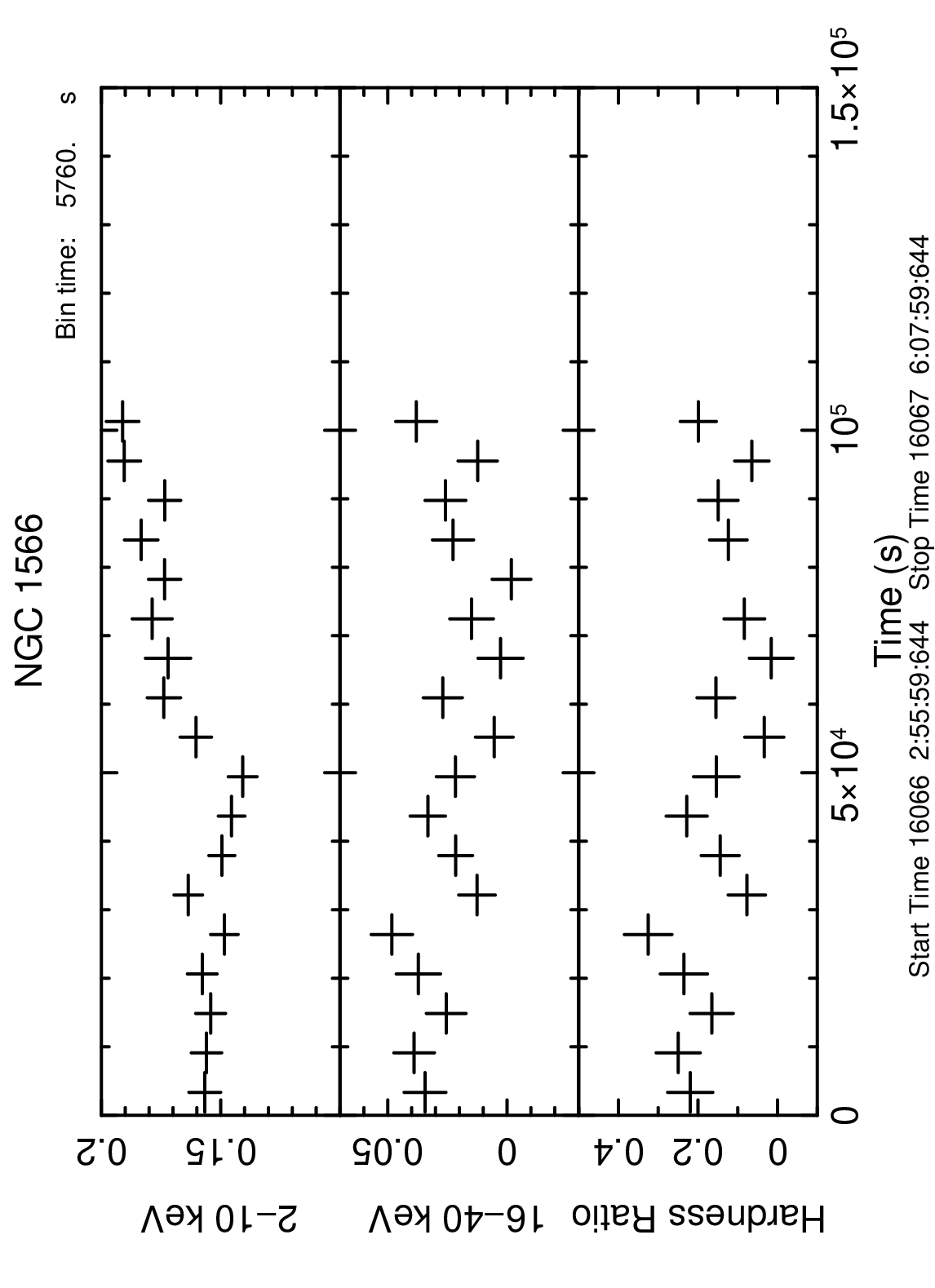}
\plotone{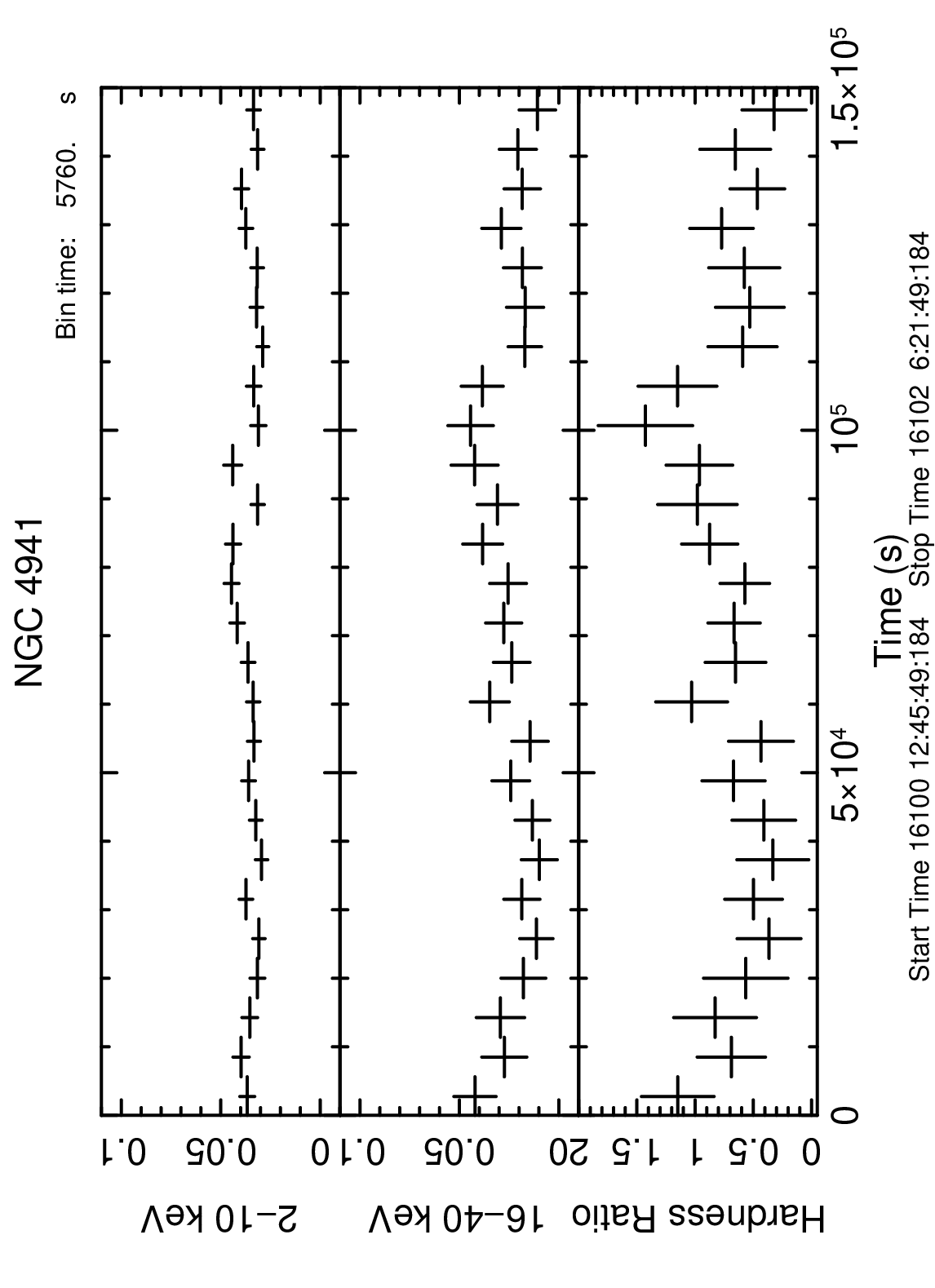}
}
\caption{
({\it Upper figure}): (a) light curve of NGC~1566 in the XIS 2--10 keV
 band ($f_{\rm 2-10}$, upper panel), that in the HXD/PIN 16--40 keV band
 ($f_{\rm 16-40}$, middle), and
 the hardness ratio between them
($f_{\rm 16-40}$/$f_{\rm 2-10}$, lower).
({\it Lower figure}): (b) same plots for NGC~4941.
}
\label{LIGHT_CURVES}
\end{center}
\end{figure}

Figure~\ref{LIGHT_CURVES} shows the background-subtracted light curves
of NGC~1566 and NGC~4941 in the 2--10 keV band (XIS0~$+$~XIS3; upper
panel), $f_{\rm 2-10}$,
and in the 16--40 keV band (PIN; middle), $f_{\rm 16-40}$.
The lower panels plot
the hardness ratio between the above two energy bands,
$f_{\rm 16-40}$/$f_{\rm 2-10}$. The bin size is set to
be 5760 sec (orbital period of {\it Suzaku}) to exclude any modulations that
depend on orbital phase.  As noticed from the figure, the 2--10 keV flux
of NGC 1566 slightly increased around $t \sim$ 50 ks since the start of
the observation, although it is not evident in the 16--40 keV band. We
find, however, no differences over statistical errors in the best-fit
spectral parameters of NGC 1566 except for the flux normalization
between the first and second half of the observation. The XIS and
PIN light curves of NGC 4941 suggest only weak flux variation. Thus, we
analyze the time-averaged spectra of {\it Suzaku} for both targets.

%%%%%%%%%%%%%%%%%%%
%   LIGHT CURVES  %
%%%%%%%%%%%%%%%%%%%

\section{ANALYSIS AND RESULTS}\label{ANALYSIS_RESULT}

\subsection{Broad-Band Spectral Analysis}\label{BROAD_ANALYSIS}\label{BROADA}

\begin{figure}
\begin{center}
\epsscale{0.9}
\rotatebox{-90}{
\plotone{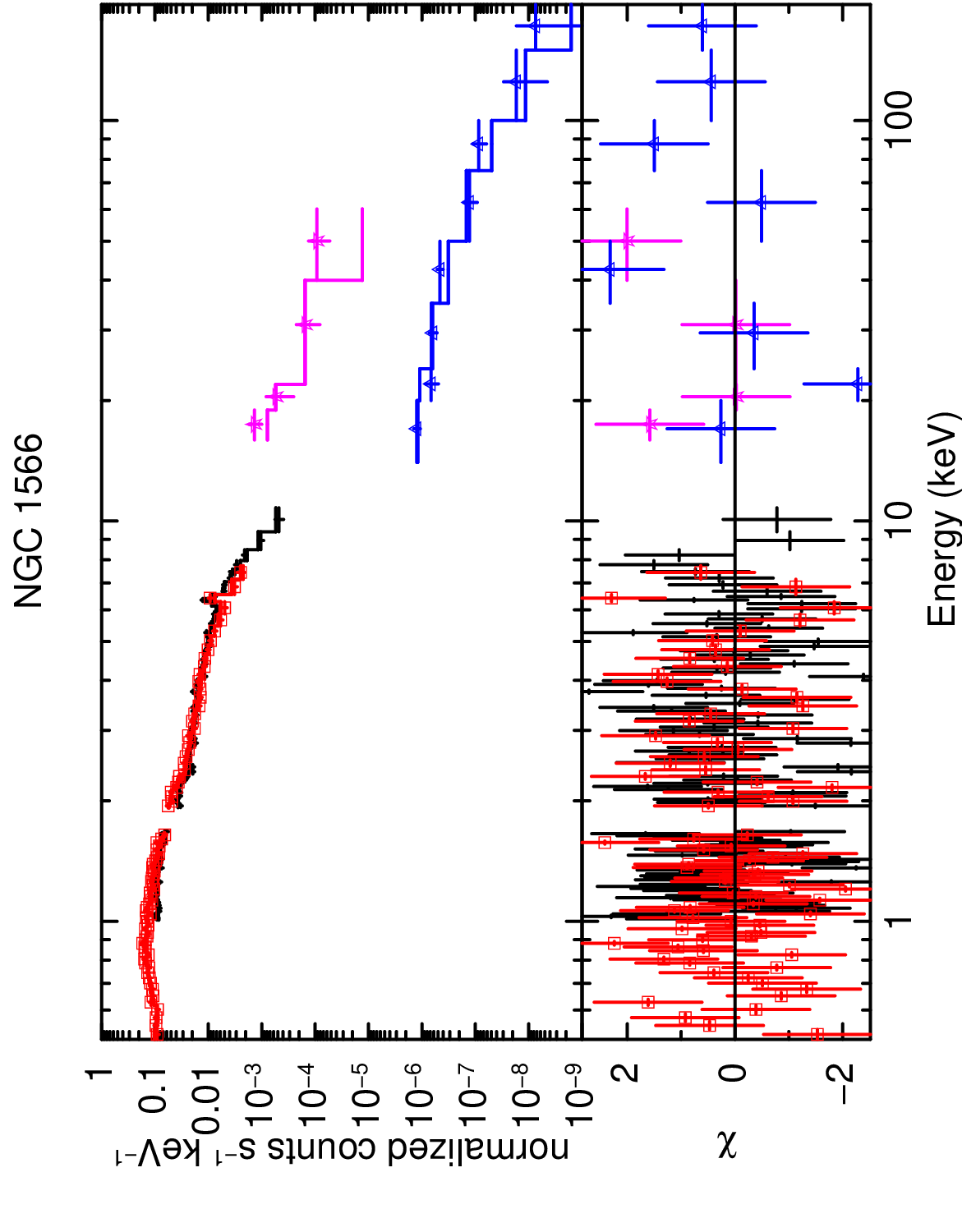}
\plotone{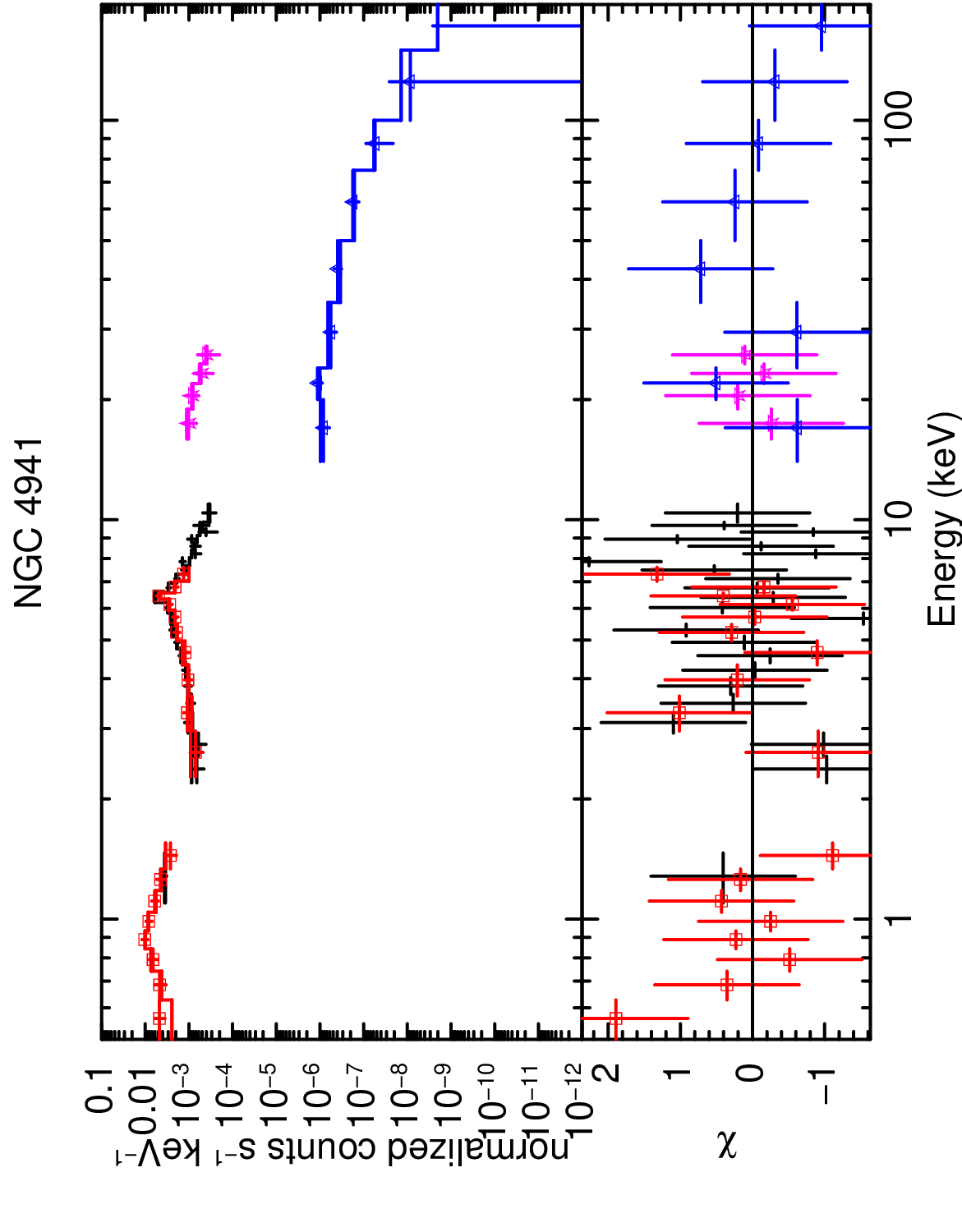}
}
\caption{
({\it Upper Figure}): (a) {\it Suzaku} and {\it Swift}/BAT spectra of NGC 1566
 (crosses). Black, red (square), magenta (star), and blue (triangle) 
 represent the FI-XISs, BI-XIS,
 HXD/PIN, and {\it Swift}/BAT spectra. The {\it Suzaku} spectra are folded
 with the energy response. The best-fit model is plotted by the solid
 curve and fitting residuals in units of $\chi$ are shown in the lower
 panel.
({\it Lower Figure}): (b) The same plots for NGC 4941.
}
\label{FOLDED_SPEC}
\end{center}
\end{figure}

For spectral analysis, we use the data of FI-XISs (XIS0 $+$ XIS3), BI-XIS
(XIS1), and HXD/PIN in the energy band of 1--12 keV, 0.5--8 keV, and
16--60 keV (16--30 keV for NGC 4941), respectively, where sufficient
signal-to-noise ratio is achieved. To obtain better constraint on the
hard X-ray spectra, we include the {\it Swift}/BAT spectra in the
14--195 keV band averaged over 70 months \citep{Bau12}. The 1.7--1.9
keV band in the XIS spectra is excluded to avoid systematic
uncertainties in the energy response around the Si K-edge region. The
{\it Suzaku} spectra folded with the responses are plotted in the
upper panels of Figure~\ref{FOLDED_SPEC}. The {\it Swift}/BAT spectra
in the photon flux unit are also shown there. As noticed, the spectrum
of NGC 1566 is essentially unabsorbed, while that of NGC 4941 is
subject to heavy absorption. Conspicuous iron-K$\alpha$ emission
lines are noticed at the rest-frame 6.4 keV in both targets.

We simultaneously fit the {\it Suzaku} and {\it Swift}/BAT spectra, which cover the
wide energy range of 0.5--195 keV altogether. To absorb the
cross-calibration error in absolute fluxes between the XISs and the HXD,
the relative normalization of the HXD/PIN to FI-XISs is fixed to be 1.16
according to the result based on the Crab Nebula observations
\citep{Mae08}, while that between BI-XIS and FI-XISs is set free 
because they cover the similar energy bands and even a small 
systematic error in their relative flux calibration would affect 
the fit significantly.
We do not apply such correction for instrumental calibration between the
{\it Suzaku} and {\it Swift}/BAT data, although time variability between the two
periods is taken into account, as detailed below.

%%%%%%%%%%%%%%%%%%%%
%  FOLDED SPECTRA  %
%%%%%%%%%%%%%%%%%%%%

We follow previous work by \citet{Taz13}, who 
constrain 
the torus structures of two very luminous radio-loud AGNs, 3C 206 and PKS 0707--35, 
by applying a physically motivated spectral model to 
the data of {\it Suzaku} and {\it Swift}/BAT.
In this paper we basically adopt the same model as in \citet{Taz13}, which is
composed of four components for the nucleus emission:
(1) the primary transmitted component from the nucleus, (2)
reflection component from the accretion disk, (3) that from the torus,
and (4) scattered component from surrounding gas in the case of
absorbed AGN (NGC 4941). In addition, we also consider emission from
an optically-thin thermal plasma in the host galaxy, which is often
observed from LLAGNs as ``soft excess'' below $\sim$1 keV
\citep{Tera02}. It is modelled with the {\bf apec} model \citep{Smi01}
on XSPEC. We approximate the shape of the primary continuum with an
exponential cutoff power law model, $E^{-\Gamma} \exp (-E/E_{\rm
cut})$. Since it is not possible to set a meaningful limit on $E_{\rm
cut}$ from our data, we fix it at 360 keV
for consistency with 
the torus model by \citet{Ike09}. Even if we instead set $E_{\rm cut}$
at 300 keV, an averaged value in nearby AGNs reported by \citet{Dad08},
the results of our spectral fits are little affected.

Time variability of the direct component flux is expected between the
short (2 days) {\it Suzaku} observation and the long (70 months) {\it
Swift}/BAT observations. To reflect it, we introduce a normalization
factor of {\it Suzaku} (FI-XISs) relative to the {\it Swift}/BAT, $Norm_{\rm
XIS}$, as a free parameter, assuming that the continuum shapes (i.e.,
$\Gamma$ and $E_{\rm cut}$) are constant. The same factor is also
applied to the reflection component from the accretion disk, while we
do not apply it to that from the torus, the scattered component, and
the thin thermal emission, assuming that their fluxes do not change
over years because of their large spatial scales.

For calculation of the reflection components, we utilize the {\bf
pexmon} model \citep{Nan07}, which consists of the continuum of the
{\bf pexrav} model \citep{Mag95} from cold matter and fluorescence
lines of iron-K$\alpha$, iron-K$\beta$, and nickel-K$\alpha$
self-consistently calculated with the continuum. The only free
parameter of the model is the reflection strength $R \equiv
\Omega/2\pi$, where $\Omega$ is the solid angle of the reflector
covering the X-ray source. As for the reflection component from the
accretion disk, we apply blurring due to Kepler motion and
relativistic effects around a non-rotating black hole with the
convolution model {\bf rdblur}.

One important goal is to constrain the torus structure through its
reflection component $R_{\rm torus}$. Since the two reflection
components from the torus and accretion disk are strongly coupled with
each other in spectral fit, we need to fix the reflection strength
from the accretion disk on the basis of reasonable
assumptions. Following \citet{Taz13}, we estimate a value of $R$ from
the disk ($R_{\rm disk}$) by combining a theoretical equivalent width
of an iron-K$\alpha$ line given by \citet{Ge91} and the {\bf pexmon}
reflection code. \citet{Ge91} calculates the iron-K$\alpha$ equivalent
width from an optically thick plane irradiated by a primary X-ray
source above it for three parameters, (i) ratio between the height of
the source ($h_{\rm s}$) and the inner radius of the disk ($r_{\rm
in}$) , (ii) inclination, and (iii) photon index. We assume $h_{\rm s}
= 10 r_{\rm g}$ ($r_{\rm g} \equiv GM/c^2$ is the gravitational radius
where $G$, $M$, and $c$ is the gravitational constant, black hole
mass, and light velocity, respectively).  Some works suggest that the
accretion disk of LLAGNs with $L_{2-10} < 10^{42}$ ergs s$^{-1}$ is
not extending down to the innermost stable circular orbit, but likely
to be truncated at much larger radii, $\sim 10^2 r_{\rm g}$ 
(e.g., \citealt{Qua99} for M81 and NGC 4579, \citealt{Ptak04} for
NGC 3998, \citealt{Nem06} for NGC 1097).  We thus adopt the inner
radius to be $100 r_{\rm g}$ in our models.  The inclination is
assumed to be 30$^\circ$ for NGC 1566 (unobscured AGN) and 70$^\circ$
for NGC 4941 (obscured AGN). We consider a range of photon index of
$\Gamma=$1.9--2.1 as obtained from our data. The expected equivalent
width of the iron-K$\alpha$ line is then calculated to be 15--21 eV
(NGC 1566) and 14--16 eV (NGC 4941). We find that the corresponding
reflection strength of the {\bf pexmon} model that produces these
values becomes $R_{\rm disk} \simeq 0.1$ in both targets.
We confirm that the spectral parameters obtained from 
the broad band fits are not changed over the 90\% confidence
errors even if we assume different inclination angles within a 
range of 20$^\circ$-60$^\circ$ for NGC 1566 and 
30$^\circ$-80$^\circ$ for NGC 4941.
Effects by changing $r_{\rm in}$ or $h_{\rm s}$ (and hence $R_{\rm disk}$)
will be examined in the following subsections, which do not affect
our conclusions, either.

\subsubsection{NGC 1566}

We first apply a simple model without any intrinsic absorption
composed of the transmitted emission and torus reflection component to
the spectra of NGC 1566, which is optically classified as a type 1.5
AGN. The latter is required to explain the strong narrow iron-K line.
The model is represented as
\begin{itemize}
\item {\bf \scriptsize \hspace{-2mm} constant*zpowerlw*zhighect+pexmon}
\end{itemize}
in XSPEC terminology. The {\bf constant} term takes into account flux
variability of the direct component between the
observation epochs of {\it Suzaku} and {\it Swift}/BAT, and the power
law normalizations in the two terms are tied together.
The inclination angle in the {\bf pexmon} model is set to 60$^\circ$
as a representative value, 
since the code assumes reflection from a half-infinite plane, which 
is too simple to be applied for a complex torus-like geometry. 
Changing it to 30$^\circ$, however, does not
affect our fitting results within the errors.
Since a majority of the torus reflection in type-1 AGNs
can arise from its inner wall without self obscuration (see e.g., 
\citealt{Ike09} for an example of torus geometry), 
we do not apply any absorption to this component.
The resultant chi-squared value is not acceptable, $\chi^2/d.o.f. =
279.6/186$.

To improve the fit,
we add a partial covering model to the first component,
which is known to sometimes give a good description of AGN spectra.
The likely interpretation is that there is patchy material in the 
line of sight. This model composition is represented as 
\begin{itemize}
\item {\bf \scriptsize \hspace{-2mm} constant*zpcfabs*zpowerlw*zhighect+pexmon},
\end{itemize}
where {\bf zpcfabs} represents an absorption by cold matter with a column
density $N_{\rm H}$ and a covering fraction of $f$. The addition of 
{\bf zpcfabs} significantly improves the fit, yielding a chi-squared 
value of $\chi^2/d.o.f. = 220.4/184$. Note that the fit is considerably 
worse when a full covering model (i.e., $f=1$) is applied
($\chi^2/d.o.f. = 279.7/185$).
Next, we add emission from an optically-thin thermal plasma.
This model, represented as
\begin{itemize}
\item {\bf \scriptsize \hspace{-2mm} constant*zpcfabs*zpowerlw*zhighect+pexmon+apec},
\end{itemize}
better explains the spectra with $\chi^2/d.o.f. = 214.0/182$, 
and the improvement 
is confirmed at $>95\%$ confidence level on the basis of an F-test.
We obtain $\Gamma = 2.03^{+0.10}_{-0.09}$, 
$R_{\rm torus} = 0.45^{+0.13}_{-0.10}$, and the variability factor 
({\bf constant}) $Norm_{\rm XIS} = 0.26^{+0.06}_{-0.05}$.

%%%%%%%%%%%%%%%%%%%%%%%%%%%%%%%
%   UNFOLDED SPECTRA NGC ---- %
%%%%%%%%%%%%%%%%%%%%%%%%%%%%%%%
\begin{figure}
\begin{center}
\epsscale{0.9}
\rotatebox{-90}{
\plotone{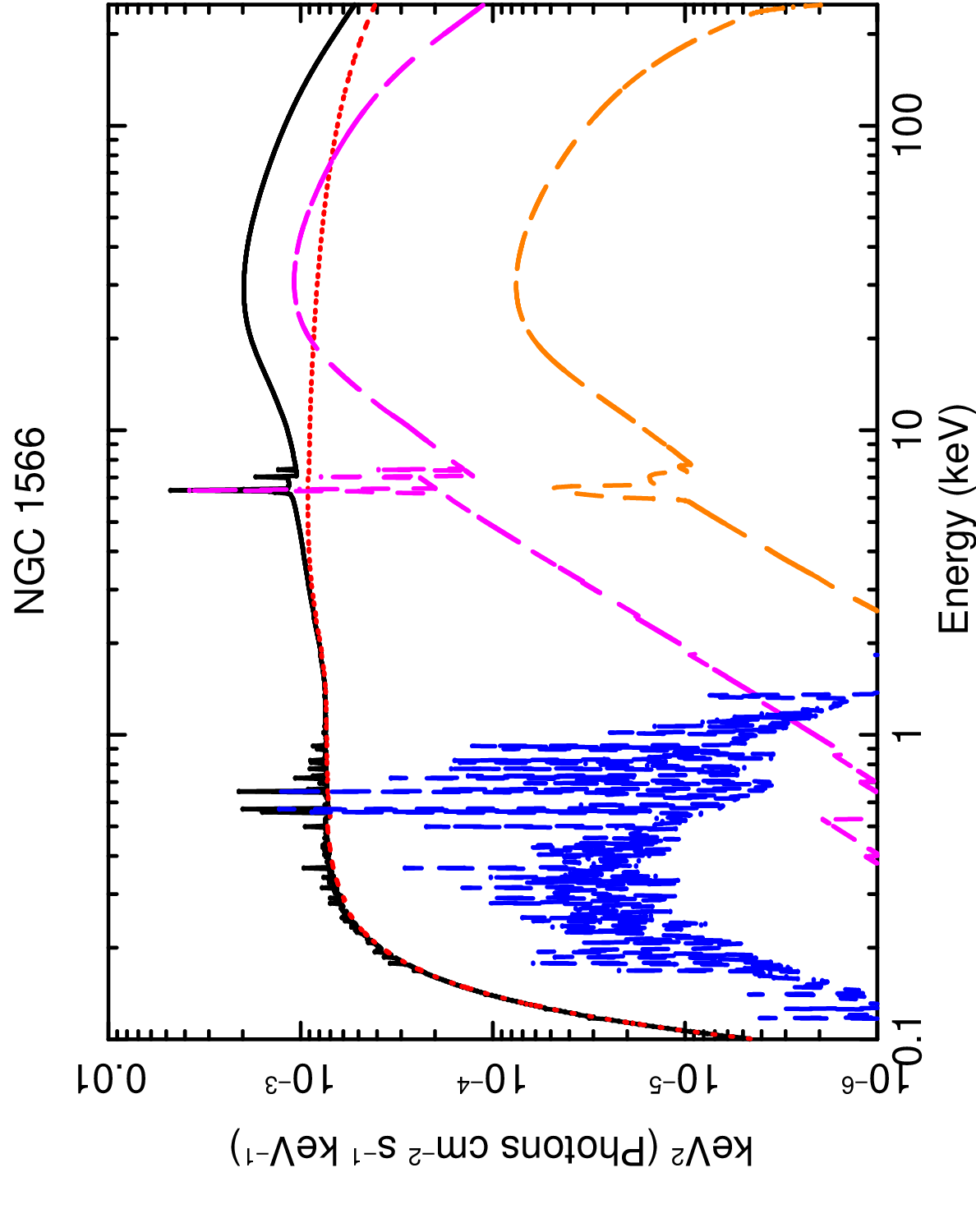}
\plotone{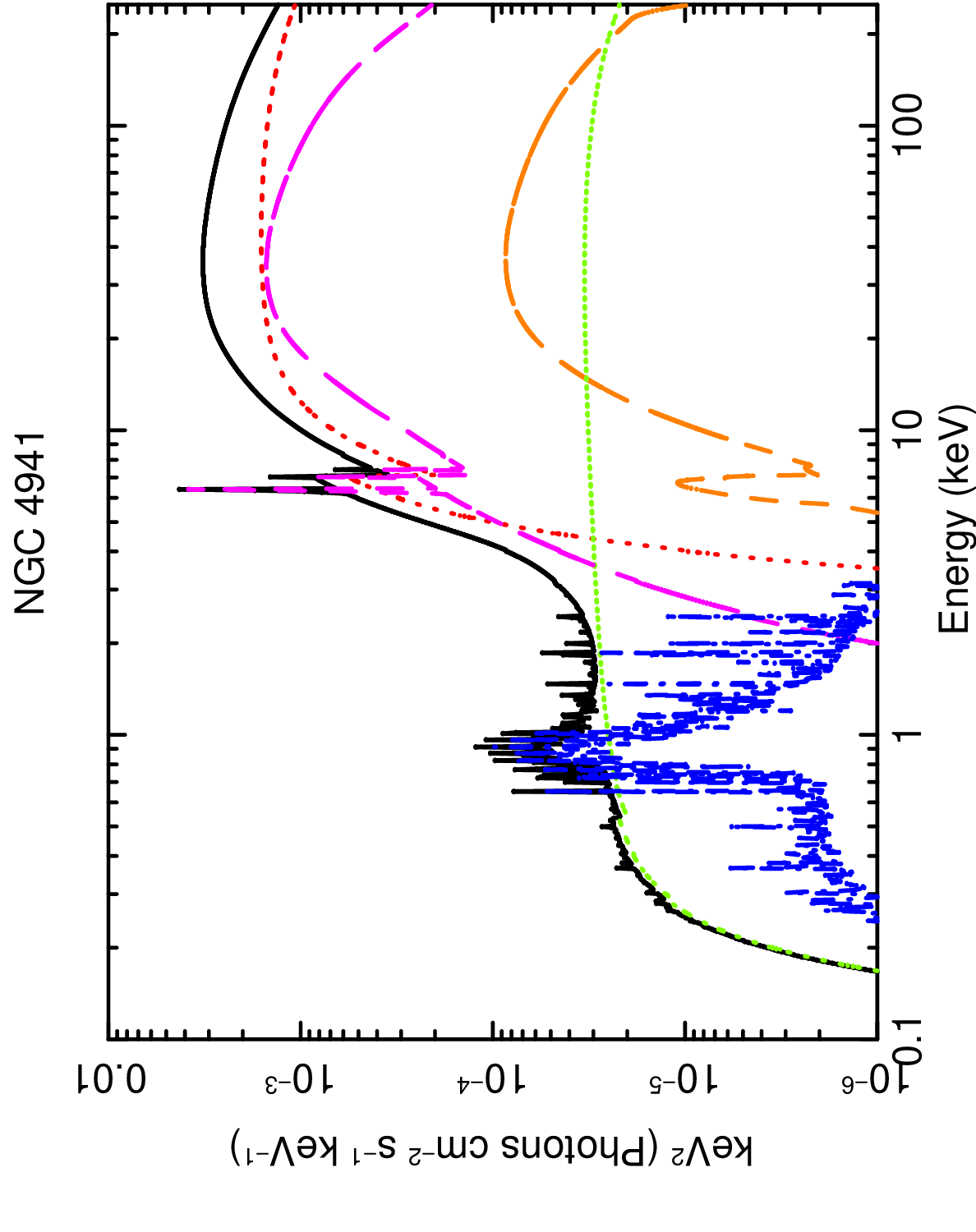}
}
\caption{
({\it Upper Figure}): (a) unfolded spectra of NGC~1566 in units of $E I_E$
 where $I_E$ is the energy flux. The lines represent
the total component (solid black), 
the transmitted component (dotted red),
the reflection component from the accretion disk (dashed orange), 
the reflection component from the torus (dashed magenta), 
the scattered component (dotted green), and 
the emission from optically-thin thermal plasma (dashed-dotted blue).
({\it Lower Figure}) : (b) same plots for NGC 4941.
}
\label{UNFOLDED_SPEC}
\end{center}
\end{figure}

Finally, as a physically motivated model introduced by \citet{Taz13},
we further include the reflection
component from the accretion disk into the above model. This ``final'' model 
is expressed as
\begin{itemize}
\item {\bf \scriptsize  \hspace{-2mm} constant*zpcfabs*(zpowerlw*zhighect+rdblur*pexmon)\\+pexmon+apec},
\end{itemize}
where the second term represents the
disk reflection component whose intensity 
is synchronized with that of the direct component (the first term).
The reflection strength from the accretion disk is fixed at $R_{\rm
disk} = 0.1$ (see above), while that from the torus $R_{\rm torus}$ is 
still left to be a free parameter. 
Note that, unlike $R_{\rm disk}$, $R_{\rm
torus}$ is defined relative to the flux determined with {\it Swift}/BAT, not
to that of {\it Suzaku}. 
The inclination in the {\bf pexmon} model is set to
be 30$^\circ$ for the disk reflection.
In the {\bf rdblur} model, we fix the inner and outer radii
at $r_{\rm in} = 100 r_{\rm g}$ and $r_{\rm out} = 10^5 r_{\rm g}$,
respectively, with an emissivity index of $\beta=-3$. 
This outer radius is appropriate
if the disk is connected to the pc scale torus, although the choice
of $r_{\rm out} > 10^3 r_{\rm g}$ does not affect the fit at all.
The index is based on a simple assumption 
of an irradiated disk by a source with a scale height above it
 \citep[see e.g.,][]{Rey97}.
This model results in a chi-squared value of $\chi^2/d.o.f. = 215.5/182$.
The fitted parameters are summarized in Table \ref{BESTFIT_PARA}.
We obtain the photon index $\Gamma = 2.04^{+0.10}_{-0.09}$ and the
reflection strength from the torus $R_{\rm torus} = 0.45^{+0.13}_{-0.10}$. 
The variability factor ({\bf constant}) 
becomes $Norm_{\rm XIS} = 0.23^{+0.06}_{-0.05}$, 
indicating the flux of the direct component
during the {\it Suzaku} observation was much lower compared with the averaged
level as determined with {\it Swift}/BAT. 
The best-fit model folded with the
responses is overplotted to the data in the upper panels of Figure~\ref{FOLDED_SPEC},
and the residuals in units of $\chi$ are shown in the lower panels.
Figure~\ref{UNFOLDED_SPEC} shows the unfolded spectrum in units of $E I_E$ ($I_E$ is the
energy flux at $E$), where different components are plotted separately.

%%%%%%%%%%%%%%%%%%%%%%%%%
%   BESTFIT PARAMETERS  %
%%%%%%%%%%%%%%%%%%%%%%%%%
\begin{deluxetable*}{cccc}
\tabletypesize{\footnotesize}
\tablecaption{The Parameters from the Best-Fit Model\label{BESTFIT_PARA}}
\tablewidth{0pt}
\tablehead{\colhead{           } &\colhead{           } & \colhead{NGC~1566} & \colhead{NGC~4941}}
\startdata
(1) & $N^{\rm{Gal}}_{\rm{H}} $ $(10^{20}$ $ {\rm cm}^{-2})$ & $0.861^a$ & $2.17^a$ \\
(2) & $N_{\rm{H1}} $ $(10^{22}$ $ {\rm cm}^{-2})$& $3.04^{+1.08}_{-0.82}$ & $73^{+19}_{-15}$ \\
(3) & $N_{\rm{H2}} $ $(10^{22}$ $ {\rm cm}^{-2})$& - & $5.1^{+7.2}_{-4.1}$ \\
(4) & $ f_{\rm c} $ & $0.26\pm0.10 $ & -  \\
(5) & $ \theta_{\rm inc} ({\rm degree})$ & $30^a $ & $70^a$  \\
(6) & $ r_{\rm in} (r_{\rm g})$ & $100^a $ & $100^a$  \\
(7) & $ R_{\rm disk} $ & $0.1^a $ & $0.1^a$  \\
(8) & Norm$_{\rm{XIS}} $ & $0.23^{+0.06}_{-0.05}$ & $0.48^{+0.35}_{-0.18}$ \\
(9) & $\Gamma$ & $2.04^{+0.10}_{-0.09}$ & $1.91^{+0.28}_{-0.23}$ \\
(10) & $f_{\rm{scat}}\Hp (\%)$ & - & $0.99^{+1.76}_{-0.63}$ \\
(11) & $R_{\rm torus}$ & $0.45^{+0.13}_{-0.10}$ & $0.64^{+0.69}_{-0.27}$ \\
(12) & $kT\Hp (\rm{keV})$ & $0.20^{+0.10}_{-0.06}$ & $0.86\pm 0.16$ \\
(13) & $F_{0.5-2}~{(\rm ergs~cm^{-2}~s^{-1})}$ & $2.30\times 10^{-12}$  & $3.12\times 10^{-14}$ \\
(14) & $F_{2-10}~{(\rm ergs~cm^{-2}~s^{-1})}$ & $2.79\times 10^{-12}$ & $4.25\times 10^{-12}$ \\
(15) & $F_{10-50}~{(\rm ergs~cm^{-2}~s^{-1})}$ & $5.14\times 10^{-12}$ & $8.40\times 10^{-12}$ \\
(16) & $F_{\rm{apec}:0.5-2}~{(\rm ergs~cm^{-2}~s^{-1})}$ & $8.08\times 10^{-14}$  & $3.05\times 10^{-14}$ \\
(17) & $L_{2-10}~{(\rm ergs~s^{-1})}$ & $9.12\times 10^{40}$ & $1.98\times 10^{41}$ \\
   & $\chi^2/d.o.f.$ & $215.47/182$ & $28.72/43$ \\ 
\enddata
\tablecomments{
(1) The hydrogen column density of Galactic absorption along the line of sight by \citet{Kal05}. \\
(2) The hydrogen column density for the partial covering model in the fraction of (4) in NGC~1566 and 
for the direct component in NGC~4941.\\
(3) The hydrogen column density for the reflection component from the dust torus.\\
(4) The covering fraction with the hydrogen column density of (2) in NGC~1566.\\
(5) The inclination angle to the accretion disk.\\
(6) The inner truncated radius of the accretion disk.\\
(7) The relative reflection strength ($R\equiv \Omega/2\pi$) of the 
accretion disk  to the flux measured with {\it Swift}/BAT.\\
(8) The normalization ratio of the flux measured with {\it Suzaku} relative to the one with {\it Swift}/BAT.\\
(9) The photon index of the powerlaw model.\\
(10) The scattering fraction relative to the flux measured with {\it Swift}/BAT. \\
(11) The relative reflection strength ($R\equiv \Omega/2\pi$) of the 
dust torus to the flux measured  with {\it Swift}/BAT.\\
(12) The temperature $kT$ (keV) of the {\bf apec} model.\\
(13) The observed flux with BI-XIS of {\it Suzaku}   in the 0.5--2 keV band. \\ 
(14) The observed flux with FI-XISs of {\it Suzaku} in the  2--10 keV band. \\ 
(15) The observed flux with HXD/PIN of {\it Suzaku} in the 10--50 keV band. \\ 
(16) The flux of the {\bf apec} model with the correction for the absorption with BI-XIS of {\it Suzaku} in the 0.5--2 keV band. \\ 
(17) The intrinsic luminosity obtained with FI-XISs of {\it Suzaku} corrected for the absorption in the 2--10 keV band.\\ 
}
\tablenotetext{a}{The parameters are fixed in the fitting.}
\end{deluxetable*}

We note that the presence of the disk reflection
component is neither required nor rejected at 90\% confidence limits
from our data, by comparison with the model without it, which gives a
slightly better chi-squared value (214.0). Nevertheless, we
adopt it in our paper as the best physically-motived model, because
the reflection component from a truncated standard disk at $\sim 100
r_{\rm g}$, as suggested from previous studies of LLAGNs (\citealt{Qua99}, 
\citealt{Ptak04}),
should be expected even though it is weak. Here we derive an upper
limit for the strength of the disk reflection $R_{\rm disk}$ from the
data. By adopting different values for the inner radius of $r_{\rm
in} = 10 r_{\rm g}$, $30 r_{\rm g}$, and $50 r_{\rm g}$, which
corresponds to $R_{\rm disk} = 0.6$, 0.3, and 0.2 for the source scale
height of $h_{\rm s} = 10 r_{\rm g}$ (see section \ref{BROADA}), we obtain
chi-squared values of 224.5, 219.5, 217.5, respectively, from the
broad band spectral fit. Thus, the data constrain $R_{\rm disk} <
0.2$ at a 90\% confidence limit, which is well consistent with our best
model, $R_{\rm disk} = 0.1$. If we instead assume $h_{\rm s} = 100
r_{\rm g}$, then $R_{\rm disk} = 0.6$ is expected for $r_{\rm in} =
100 r_{\rm g}$ and the fit results in a worse chi-squared value of
224.3. This implies that the smaller scale height, 
$h_{\rm s} = 10 r_{\rm g}$, is more reasonable unless the disk is
truncated at a very large radius like $r_{\rm in} > 10^3 r_{\rm g}$.
In any case, we
confirm that our conclusion on the torus structure based on the 
iron-K$\alpha$ line analysis described in section~\ref{TORUS}
is not affected by these uncertainties on the disk-reflection component,
whose contribution to the total spectrum is rather small.

\subsubsection{NGC 4941}

We first apply the following model for the broad band spectra of NGC 4941, 
which is a type-2 AGN and shows clear evidence for an intrinsic absorption:
\begin{itemize}
\item {\bf \scriptsize  \hspace{-2mm} constant*zphabs*zhighect*zpowerlw+zphabs*pexmon
\\+constant*zhighect*zpowerlw}
\end{itemize}
The first and second terms are the transmitted and 
disk reflection components, respectively. Here
we consider absorptions separately for the 
two components, with $N_{\rm H1}$
for the former and with $N_{\rm H2}$ for the latter, because
the torus reflection comes from a large volume
and hence could well be subject to a different 
absorption in average from that for the transmitted emission.
The third term represents a scattering component 
by a gas surrounding the nucleus. It is assumed to have 
the same photon index $\Gamma$ as the transmitted
component with a relative normalization of $f_{\rm scat}$ (to the
{\it Swift}/BAT flux). 
%We obtain the best-fit photon index of 
%$\Gamma=2.19^{+0.26}_{-0.24}$, $N_{\rm H1}**=(88^{+33}_{-20})\times10^{22}$
%cm$^{-2}$, $N_{\rm H2}**=(8.4^{+7.0}_{-3.8})\times10^{22}$ cm$^{-2}$, 
%$R_{\rm torus} = 0.44^{+0.31}_{-0.14}$, 
%$f_{\rm scat}=(0.49^{+0.73}_{-0.30})$\%, and 
%$Norm_{\rm XIS} = 0.35^{+0.17}_{-0.12}$
Even though the chi-squared is already 
acceptable with this model ($\chi^2/d.o.f. = 44.3/45$), 
we add emission from a thin thermal plasma to the above model, which is
often observed in type-2 AGNs \citep[e.g.,][]{Tur97} including LLAGNs \citep{Tera02}. 
The model composition is then expressed as
\begin{itemize}
\item {\bf \scriptsize \hspace{-2mm} constant*zphabs*zhighect*zpowerlw+zphabs*pexmon\\
+constant*zhighect*zpowerlw+apec}.
\end{itemize}
The addition of the thin thermal emission significantly improves the
fit, reducing the chi-squared value to $\chi^2/d.o.f. = 28.7/43$. 
We obtain $\Gamma=1.91^{+0.28}_{-0.23}$, 
$N_{\rm H1}=(74^{+19}_{-15})\times10^{22}$cm$^{-2}$, 
$N_{\rm H2}=(5.1^{+7.1}_{-4.1})\times10^{22}$ cm$^{-2}$, 
$R_{\rm torus} = 0.62^{+0.67}_{-0.26}$, 
$f_{\rm scat}=(0.95^{+1.70}_{-0.61})$\%, and $Norm_{\rm XIS} = 0.47^{+0.34}_{-0.18}$.

In the same way as the spectral analysis of NGC 1566,  
we finally fit the spectra with the model including the disk reflection component:
\begin{itemize}
\item {\bf \scriptsize  \hspace{-2mm} constant*zphabs(zhighect*zpowerlw+rdblur*pexmon)\\
+zphabs*pexmon+constant*zhighect*zpowerlw+apec}.
\end{itemize}
The reflection strength from the accretion disk is
fixed at $R_{\rm disk}=0.1$ with an inclination of 70$^\circ$ (see
section \ref{BROADA}), while 
that from the torus, $R_{\rm torus}$, is a free parameter 
defined relative to the {\it Swift}/BAT flux.
Again, an inclination of 60$^\circ$ is assumed for the torus reflection
component as a representative value so that we can more directly compare
the result of NGC 1566; the best-fit parameters are not significantly
changed even if when an inclination of 30$^\circ$ is adopted instead.
We find this model describes the observed spectra well
($\chi^2/d.o.f. = 28.72/43$) with the best-fit parameters of $\Gamma
= 1.91^{+0.28}_{-0.23}$, $N_{\rm H1}=(73^{+19}_{-15})\times10^{22}$
cm$^{-2}$, $N_{\rm H2}=(5.1^{+7.2}_{-4.1})\times10^{22}$ cm$^{-2}$, $R_{\rm torus} =
0.64^{+0.69}_{-0.27}$, and $f_{\rm scat}=(0.99^{+1.76}_{-0.63})$\%. We obtain 
$Norm_{\rm XIS} = 0.48^{+0.35}_{-0.18}$, which indicates that the direct component flux was
weaker by 52\% during the {\it Suzaku} observation than the 70-month average
observed with {\it Swift}/BAT.  
Unlike the case of NGC 1566, we cannot constrain the strength of the
disk reflection from our data, mainly due to the heavy obscuration of the direct
and disk-reflection components. We obtain a similar chi-squared value 
of $\chi^2/d.o.f.=28.0/43$ even when we assume $r_{\rm in} = 6 r_{\rm g}$
corresponding to $R_{\rm disk} = 0.7$ for $h_{\rm s} = 10 r_{\rm g}$,
and $\chi^2/d.o.f.=28.9/43$ when $r_{\rm in} = 100 r_{\rm g}$ and 
$h_{\rm s} = 100 r_{\rm g}$ corresponding to $R_{\rm disk} = 0.6$ are adopted.
Nevertheless, with the same argument for NGC 1566, 
 we adopt this model that includes the disk reflection with $R_{\rm
 disk}=0.1$ as the best physical model for NGC 4941.
All the parameters are listed in
Table~\ref{BESTFIT_PARA}. The best-fit folded model is plotted in
Figure~\ref{FOLDED_SPEC} with fitting residuals (lower panel), and the unfolded
spectra in units of $E I_E$ are shown in Figure~\ref{UNFOLDED_SPEC}.

\subsection{Analysis of Iron-K$\alpha$ Line in Narrow Band Spectra}\label{NARROW_ANALYSIS}

%%%%%%%%%%%%%%%%%%%%%%%%%%%%%%%%%%%
% NARROW BAND ANALYSIS FOR TYPE 1 %
%%%%%%%%%%%%%%%%%%%%%%%%%%%%%%%%%%%
\begin{figure}
\begin{center}
\epsscale{0.9}
\rotatebox{-90}{
\plotone{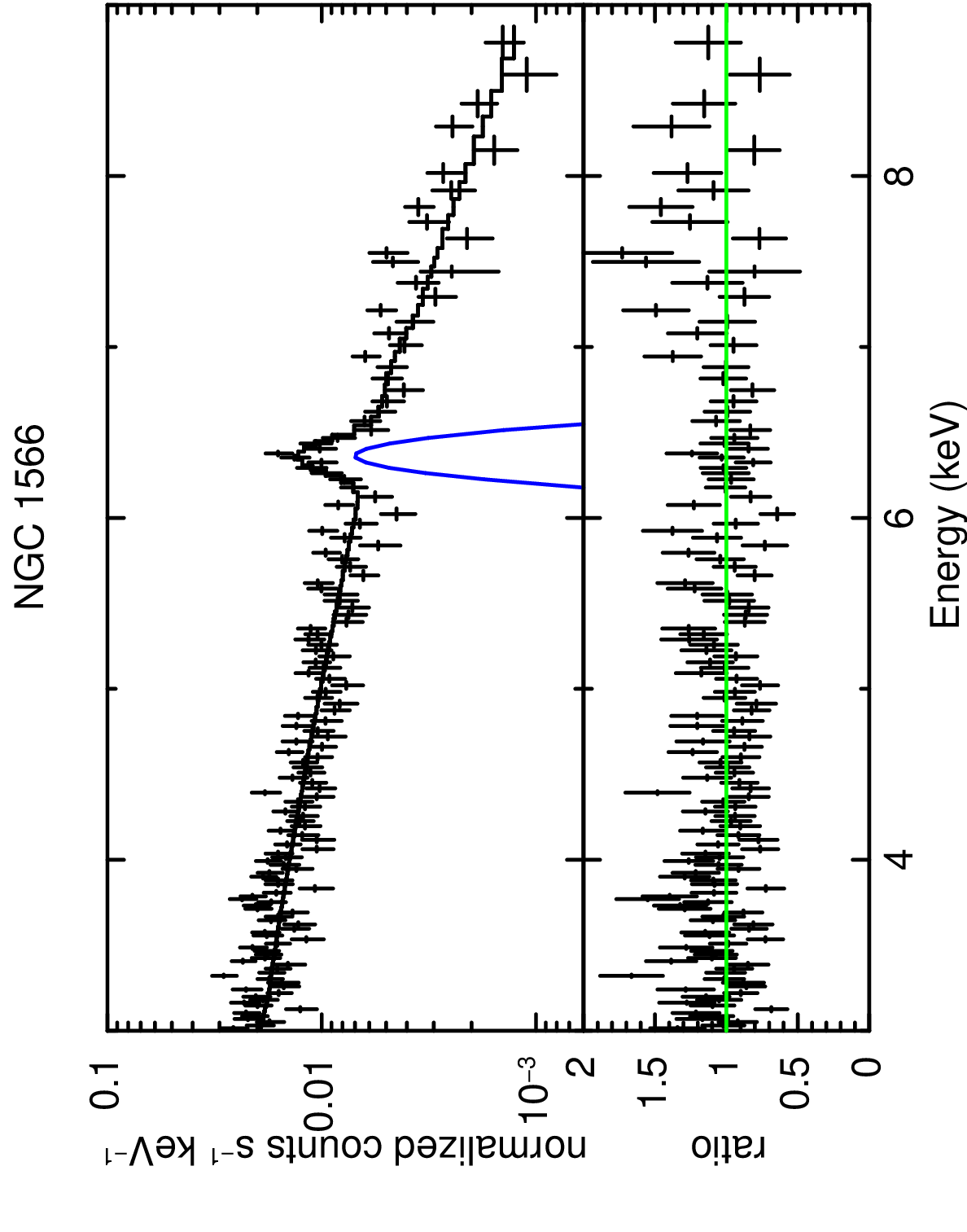}
\plotone{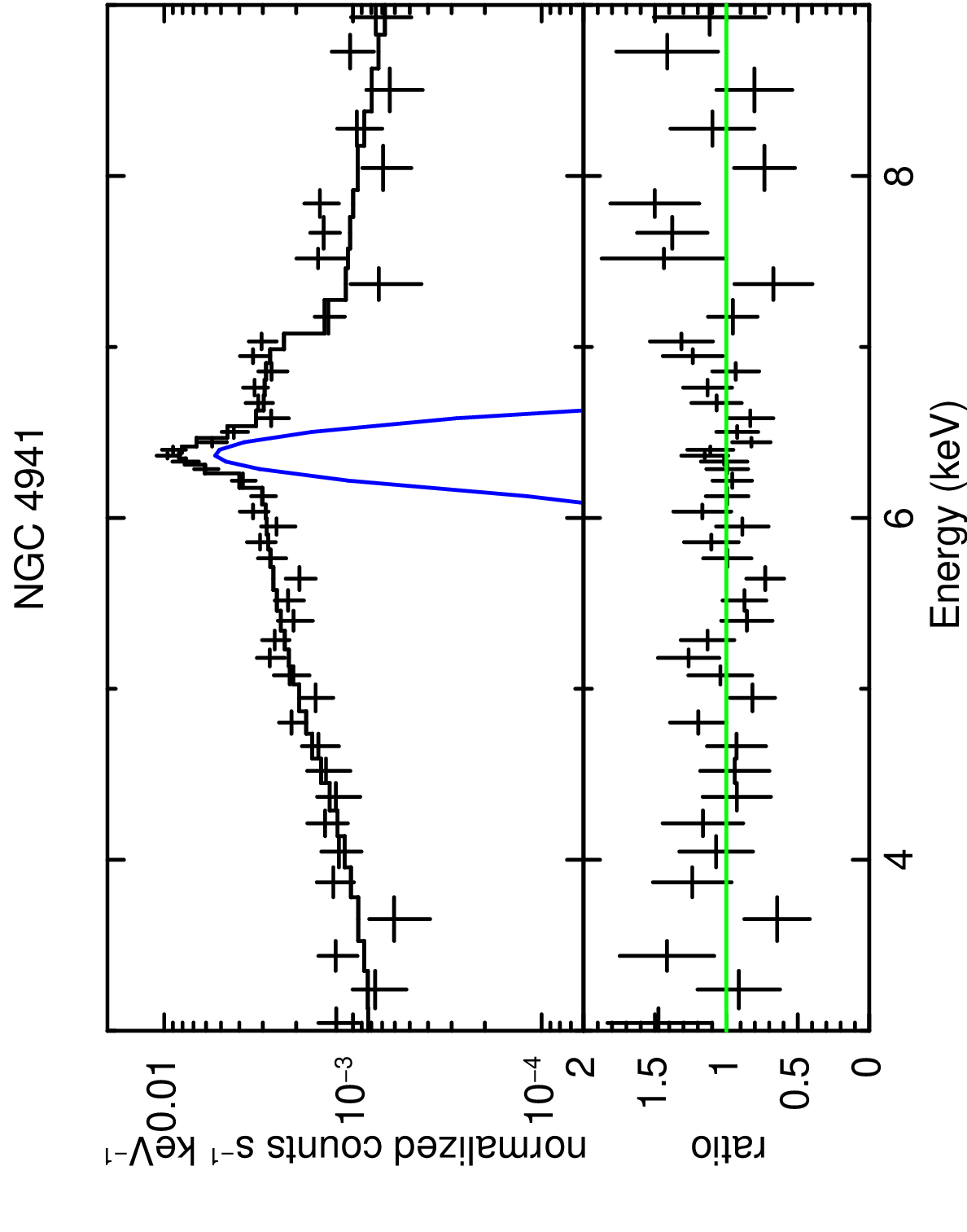}
}
\caption{
({\it Upper Figure}): (a) FI-XIS spectrum of NGC 1566 in the 3--9 keV
 band folded with the energy response (crosses). The best-fit model is
 plotted by the black solid line. The {\bf zgauss} component is shown by
 the blue solid curve.  The ratio between the data and model is plotted
 in the lower panel.
({\it Lower Figure}): (b) same plots for NGC 4941.
}
\label{NARROW_BAND}
\end{center}
\end{figure}

%%%%%%%%%%%%%%%%%%%%%%%%%
%   BESTFIT PARAMETERS  %
%%%%%%%%%%%%%%%%%%%%%%%%%
\begin{deluxetable*}{cccc}
\tabletypesize{\footnotesize}
\tablecaption{The Parameters in the Narrow-band Analysis\label{NARROW_ANALYSIS}}
\tablewidth{0pt}
\tablehead{\colhead{           } &\colhead{           } & \colhead{NGC~1566} & \colhead{NGC~4941}}
\startdata
(1) & $E_{\rm Gauss}^{\rm line} $ (keV) & $6.4^a$ & $6.4^a$ \\
(2) & $E_{\rm disk}^{\rm line} $ (keV) & $6.4^a$ & $6.4^a$ \\
(3) & $r_{\rm in}$ $(r_{\rm g}) $ & $100^a$ & $100^a$ \\
(4) & $ \theta_{\rm inc}$ $({\rm degree})$ & $30^a $ & $70^a$  \\
(5) & $N_{\rm Gauss}$ $(10^{-6}$ ${\rm photons}$ ${\rm cm}^{-2} $ ${\rm s}^{-1})$& $6.6\pm1.2$ & $6.7\pm1.2$ \\
(6) & $ {\rm EW}_{\rm Gauss}^{\rm obs}$ $({\rm eV})$ &  $240\pm40$ & $380\pm80 $   \\
(7) & $ {\rm EW}_{\rm Gauss}^{\rm cor}$ $({\rm eV})$ &  $60\pm10$ & $230\pm50$    \\
(8) & $N_{\rm disk} $ $(10^{-5}$ ${\rm photons}$ ${\rm cm}^{-2} $ ${\rm s}^{-1})$ & $<0.39$ & $<1.83$ \\
(9) & $ {\rm EW}_{\rm disk}^{\rm obs}$ $({\rm eV})$ &  $<32$ & $<150 $   \\
(10) & $ {\rm EW}_{\rm disk}^{\rm cor}$ $({\rm eV})$ &  $<39$ & $<230 $   \\
\enddata
\tablecomments{
(1) The line energy of iron-K$\alpha$ line in the {\bf zgauss} model.\\
(2) The line energy of iron-K$\alpha$ line at the rest frame in the model {\bf diskline}.\\
(3) The inner truncated radius of the accretion disk.\\
(4) The inclination angle to the accretion disk.\\
(5) The photon flux of the model {\bf zgauss} at the line energy of (1).\\
(6) The observed equivalent width of the narrow iron-K$\alpha$ line 
relative  to the total continuum observed with {\it Suzaku}.\\
(7) The equivalent width of the narrow iron-K$\alpha$ line  relative to the continuum summed over the 
direct component and the reflection component from the dust torus.\\
(8) The photon flux of the model {\bf diskline} at the line energy of (2).\\
(9) The observed equivalent width of the broad iron-K$\alpha$ line relative  to the total continuum
 observed with {\it Suzaku}.\\
(10) The equivalent width of the broad iron-K$\alpha$ line derived from the photon 
flux of (9) relative to the continuum summed over the direct 
component and the reflection component from the accretion disk.\\
}
\tablenotetext{a}{The parameters are fixed in the fitting.}
\end{deluxetable*}

To verify that the above broad-band fit results, we also perform
spectral analysis in the narrow 3--9 keV band, focusing on the iron-K$\alpha$
line features. This procedure enables us to derive observed equivalent
widths (or their upper limits) most directly from the data,
independently of the assumption of the reflection model.
Here we use the spectra of the FI-XISs in the 3--9 keV band and those of
BI-XIS in the 3--8 keV band.
Basically, the best-fit continuum model obtained by the broad-band fit
is adopted by replacing the {\bf pexmon} components with the {\bf
pexrav} model, which does not contain fluorescence lines.
Instead, we add a {\bf diskline} component \citep{Fa89} as 
the iron-K$\alpha$ emission line from the
accretion disk, and {\bf zgauss} as that from the torus. The line
energies for both components are fixed at 6.4 keV in the rest frame, and
the 1$\sigma$ line width of {\bf zgauss} is set to be 0.1 eV, which is
unresolved with the XIS. The line profile parameters in the {\bf
diskline} model are fixed at the same values as those in the {\bf
rdblur} component. Thus, only the normalizations of {\bf diskline}, {\bf
zgauss}, and {\bf zpowerlw} are set free, while the other parameters are
all fixed at their best fit values determined from the broad-band spectra.
Figure 4 shows the FI-XISs spectra in the 3--9 keV band 
folded with the energy response, where the best-fit models of {\bf zgauss}
are shown.% addition

In both targets, the {\bf diskline} components are not significantly
detected with 90\% confidence upper limits on ``observed'' equivalent
widths with respect to the total continuum ${\rm EW}^{\rm obs}_{\rm
disk} < 32$ eV (NGC~1566) and ${\rm EW}^{\rm obs}_{\rm disk} < 150$ eV
(NGC~4941). The error for NGC 4941 is larger than that for NGC 1566
because of the fainter continuum flux due to the heavy absorption and
the large inclination assumed ($70^\circ$) that produces a broader
iron-K$\alpha$ line profile. To compare these upper limits with the theoretical
predictions, we need to subtract the contribution of the
torus-reflection and scattered components from the total continuum.
Thus we obtain ``corrected'' equivalent widths of ${\rm EW}^{\rm cor}_{\rm disk} < 39$ 
eV (NGC~1566) and ${\rm EW}^{\rm cor}_{\rm disk}
< 230$ eV (NGC~4941). These upper limits are consistent with the
calculation by \citet{Ge91} under our assumptions on the geometry
between the primary source and disk ($h_{\rm s}/r_{\rm in} = 0.1$) and
the inclination, 15--21 eV for NGC~1566 and 14--16 eV for NGC~4941 (see
section~\ref{ANALYSIS_RESULT}). Hence, adopting the disk-reflection
strength of $R_{\rm disk} = 0.1$ in the broad-band spectral analysis is
justified. Note that even when we adopt a smaller inner radius of
$r_{\rm in} < 100 r_{\rm g}$ in the {\bf diskline} and {\bf rdblur}
models, the broad iron-K$\alpha$ line is not significant detected; for $r_{\rm
in}$ = $6 r_{\rm g}$, $10 r_{\rm g}$, or $30 r_{\rm g}$ we obtain an
upper limit of ${\rm EW}^{\rm cor}_{\rm disk} < 42$ eV in NGC 1566,
although the constraint is weaker in NGC 4941.

We find that the equivalent widths of {\bf zgauss} relative to the
total continuum are ${\rm EW}^{\rm obs}_{\rm Gauss} = 240\pm40$ eV and
${\rm EW}^{\rm obs}_{\rm Gauss} = 380\pm80$ eV for NGC~1566 and
NGC~4941, respectively. The observed equivalent width of a narrow
iron-K$\alpha$ line from the torus with respect to the total continuum
is subject to variability of the transmitted component, however. From
the broad band spectral analysis, we know that the fluxes during the
{\it Suzaku} observations were much fainter than those of {\it
Swift}/BAT in both targets. This leads to an overestimate of the true
(i.e., time averaged) equivalent widths of the iron-K$\alpha$ line. On
the basis of our assumption that the long-term {\it Swift}/BAT data
give the averaged continuum flux that determines the torus reflection
strength and narrow iron-K$\alpha$ line flux observed with {\it Suzaku}, we
calculate the true continuum level at 6.4 keV by increasing the
transmitted and disk-reflection components by a factor of $1/Norm_{\rm
XIS}$, and then derive ``corrected'' equivalent width values, 
${\rm EW}^{\rm cor}_{\rm Gauss}=60\pm10$ eV for NGC 1566 and 
${\rm EW}^{\rm cor}_{\rm Gauss}=230\pm50$ eV for NGC 4941.

\section{SUMMARY AND DISCUSSION}

\subsection{Summary of Results}

With {\it Suzaku} and {\it Swift}/BAT, we have obtained the broad-band
X-ray spectra covering the 0.5--195 keV band of the two LLAGNs NGC 1566
(type 1.5) and NGG 4941 (type 2) for the first time. The spectrum of NGC
1566 is found to be essentially unobscured, while that of NGC 4941 is
subject to heavy absorption. Strong iron-K$\alpha$ emission lines at the
rest-frame 6.4 keV are detected in both targets. Their spectra are well
reproduced by a physically motivated model consisting of a partially or
fully absorbed transmitted component with its reflection from the
accretion disk, a reflection component from the torus, a scattered
component (in NGC 4941), and optically-thin thermal plasma emission. The
physical parameters of the thin thermal components (temperature of $kT
\sim$ 0.2--0.9 keV and luminosities in the 0.3--2 keV band of diffuse X-ray emission)
are consistent with the origins from star-forming activities in the host galaxies 
\citep[e.g.,][]{Tul06}, although it may be partially contaminated by line emission
from a photoionized plasma powered by the AGN.
In fact, the far-infrared luminosities calculated from the {\it IRAS} 60 $\mu$m and 100
$\mu$m fluxes by using the formula of \citet{Dav92} are $2.9\times10^{43}$ 
ergs s$^{-1}$ (NGC~1566) and $4.2\times10^{42}$ 
ergs s$^{-1}$ (NGC~4941), suggesting the presence 
of significant star-forming activities.
In the analysis, we carefully take into
account time variability of the transmitted plus disk reflection
components between the {\it Suzaku} and {\it Swift}/BAT observations,
while it is assumed that the last three components are constant. During
the {\it Suzaku} observations, the flux levels of both objects were
significantly fainter than their 70 months average obtained by {\it
Swift}/BAT.

An important result is that both NGC~1566 and NGC~4941 show moderate
mount of the torus reflection, $R_{\rm torus}=0.45^{+0.13}_{-0.10}$
and $0.64^{+0.69}_{-0.27}$, respectively, relative to
the time-averaged flux of the direct component measured by
{\it Swift}/BAT. They are slightly smaller as a typical total reflection
strength observed in Seyfert galaxies, $R\sim 1$
\citep[e.g.,][]{Dad08}. The torus structure inferred from this result
is quantitatively discussed in subsection~4.3.

We compare our results on NGC 4941 with previous results obtained by
{\it ASCA} in 1996 July and 1997 January \citep{Tera02} and {\it
Beppo}SAX in 1997 January \citep{Mai98}, although much simpler
spectral models are adopted there due to the limited statistics in the
spectra. The equivalent width of the narrow iron-K$\alpha$ line is reported
to be $570\pm230$ eV ({\it ASCA}) and $1600^{+700}_{-900}$ eV ({\it
Beppo}SAX). The {\it Suzaku} result of ${\rm EW}_{\rm Gauss}^{\rm
obs}=380\pm80$ eV may be slightly smaller than these. If we compare
the iron-K$\alpha$ line flux instead of equivalent width, however, we
find they are all consistent within the statistical errors;
$1.0\pm0.4$, $1.2^{+0.5}_{-0.6}$, and
$0.67\pm0.12\times10^{-5}$ photons cm$^{-2}$ s$^{-1}$ in the {\it
ASCA}, {\it Beppo}SAX, and {\it Suzaku} observations,
respectively. This supports our hypothesis that the absolute flux of
the reflection component originating from the torus is nearly constant
as its variability is smeared out. The column density for the
transmitted component ($N_{\rm H}= 73^{+19}_{-15}\times10^{22}$ cm$^{-2}$) is
consistent with both that suggested by \citet{Tera02} and by
\citet{Mai98} within the statistical error. We obtain the photon index
$\Gamma=1.91^{+0.28}_{-0.23}$, which is consistent with a canonical
value of AGNs \citep{Nan94, Dad08}. \citet{Tera02} report a little smaller
photon index ($\Gamma=1.48^{+0.14}_{-0.15}$) but this may be affected
by the reflection component that is ignored in their model.

\subsection{Reflection Component from the Accretion Disk}

%%%%%%%%%%%%%%%%%%%%%%%%%%%%%%%
% EW 2 OPENINGANGLE for TYPE1 %
%%%%%%%%%%%%%%%%%%%%%%%%%%%%%%%
\begin{figure}
\begin{center}
\epsscale{0.9}
\rotatebox{-90}{
\plotone{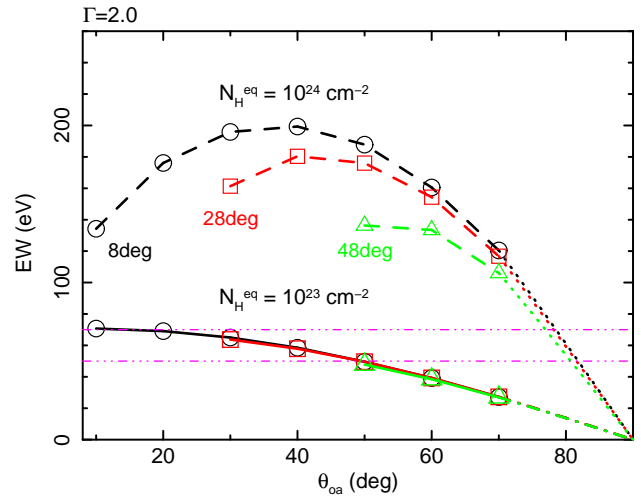}
}
\caption{
Predicted equivalent widths of iron-K$\alpha$ line for type-1 AGNs plotted as a
function of torus half-opening angle, based on the torus model by
\citet{Ike09}. 
An equatorial column density of $N_{\rm H}^{\rm eq} =
10^{23}$ cm$^{-2}$ (dashed and dotted lines) or 
$10^{24}$ cm$^{-2}$ (solid and dashed-dotted lines) 
and a photon index of 2.0 are assumed.
Several different inclinations are assumed as labeled in the figure.
The circles (black), squares (red), and triangles (green)
represent the results at the inclination angle $\theta_{\rm inc}=8^\circ$, 
28$^\circ$, and 48$^\circ$, respectively.
The 90\% confidence upper and lower limits for the corrected iron-K$\alpha$ line
equivalent width (see text) of NGC~1566 are shown by the two horizontal lines (magenta).
}
\label{EW2OPEN_TYPE1}
\end{center}
\end{figure}

%%%%%%%%%%%%%%%%%%%%%%%%%%%%%%%
% EW 2 OPENINGANGLE for TYPE2 %
%%%%%%%%%%%%%%%%%%%%%%%%%%%%%%%
\begin{figure}
\begin{center}
\epsscale{0.9}
\rotatebox{-90}{
\plotone{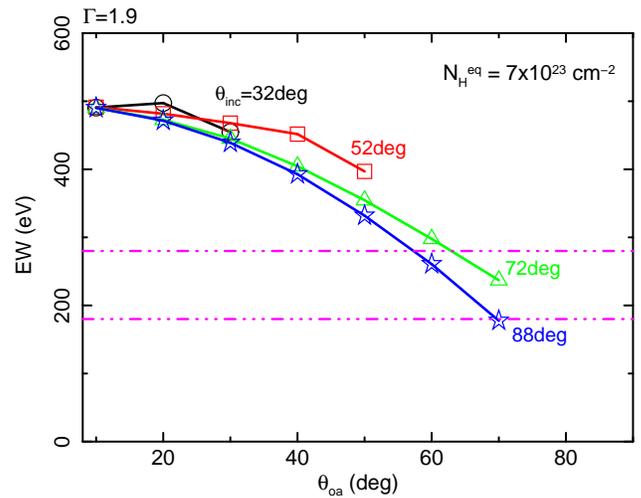}
}
\caption{
Predicted equivalent widths of iron-K$\alpha$ line for type-2 AGNs plotted as a
function of torus half-opening angle, based on the torus model by
\citet{Ike09}. An equatorial column density of $N_{\rm H}^{\rm eq} =
7\times10^{23}$ cm$^{-2}$ and a photon index of 1.9 are assumed.
Several different inclinations are assumed as labeled in the figure.
The circles (black), squares (red), triangles (green), and stars (blue)
represent the results at the inclination angle $\theta_{\rm inc}=32^\circ$, 52$^\circ$, 72$^\circ$ 
and  88$^\circ$, respectively.
The 90\% confidence upper and lower limits for the corrected iron-K$\alpha$ line equivalent
 width (see text) of NGC~4941 are shown by the two horizontal lines (magenta).
}
\label{EW2OPEN_TYPE2}
\end{center}
\end{figure}

The XIS spectra in the 3--9 keV band suggest that a broad iron-K$\alpha$
line feature arising from the inner accretion disk is very weak in NGC
1566, although it is more difficult to constrain its intensity in NGC
4941 due to the heavy absorption and high inclination. Applying a
relativistic line profile with an inner radius of $r_{\rm in} \leq 100
r_{\rm g}$, we obtain an upper limit on the corrected equivalent width
of 42 eV for NGC 1566. This result constrains the reflection strength
from a cold accretion disk to be $R_{\rm disk} < 0.3$ as defined in the
{\bf pexmon} model. In fact, when we adopt a larger value of $R_{\rm
disk} = 0.6$ expected from the case of $h_{\rm s}/r_{\rm in} = 1$ (where
$h_{\rm s}$ is the height of the irradiating source) according to
\citet{Ge91}, the broad band fit of NGC 1566 becomes significantly worse
($\chi^2/d.o.f. = 224.5/182$) compared with that for $R_{\rm} = 0.1$
($\chi^2/d.o.f. = 215.5/182$). The small reflection suggests that the
inner disk is likely to be truncated at relatively large radius,
although it is difficult to unambiguously determine the truncation
radius, which depends on the assumed corona geometry. The weak disk
reflection is consistent with 
the result by \citet{Ptak04}, who found 
$R_{\rm disk} < 0.1$ for the LLAGN NGC 3998 ($L_{2-10}\sim 10^{41}$
ergs s$^{-1}$) from the X-ray spectrum.
We do not rule out, however, the possibility that there is a
very highly ionized disk that produces almost featureless continuum
around the iron-K$\alpha$ band.

\subsection{Torus Structure}\label{TORUS}

The equivalent width of narrow iron-K$\alpha$ line from an AGN is very
useful to constrain the torus structure such as the opening angle. For
this purpose, we utilize the Monte-Carlo based numerical model by
\citet{Ike09}, which calculates absorbed direct component and the
reflected continuum with fluorescence lines from an AGN surrounded by
a torus. In this model, the geometry of the torus is close to be a
spherical shape defined by three parameters (for details see Figure~2
of \citealt{Ike09}): hydrogen column density at the equatorial plane
$N_{\rm H}^{\rm eq}$, half opening angle $\theta_{\rm oa}$, and
inclination $\theta_{\rm inc}$. Hence, $\theta_{\rm inc} < \theta_{\rm
oa}$ for type 1 AGN and $\theta_{\rm inc} > \theta_{\rm oa}$ for type
2 AGN. The incident spectrum is assumed to be a power law with an
exponential cutoff at 360 keV. As done in \citet{Taz13}, we plot
the predicted equivalent width of the iron-K$\alpha$ line as a function of
half opening angle with several different inclinations in
Figures~\ref{EW2OPEN_TYPE1} for type-1 AGNs and Figures~\ref{EW2OPEN_TYPE2}
for type-2 AGNs. 
We consider two cases of $N_{\rm H}^{\rm eq} = 10^{23}$ cm$^{-2}$ or $10^{24}$
cm$^{-2}$ with $\Gamma=2.0$ (the best fit value of NGC 1566) in
Figure~\ref{EW2OPEN_TYPE1}, while $N_{\rm H}^{\rm eq} = 7\times10^{23}$
cm$^{-2}$ and $\Gamma=1.9$, the best fit line-of-sight column density
and photon index of NGC 4941, are assumed in
Figure~\ref{EW2OPEN_TYPE2}.
Above $\theta_{\rm oa} > 70^\circ$ where the Ikeda model is not available,
we extrapolate the data by assuming that the equivalent width is
proportional to the volume of the torus.

The dashed horizontal lines (magenta) in Figures~\ref{EW2OPEN_TYPE1} 
and \ref{EW2OPEN_TYPE2} represent
the error region of the corrected equivalent width ${\rm EW}_{\rm
Gauss}^{\rm cor}$ for NGC 1566 and NGC 4941, respectively. 
Although the result of NGC 1566 
does not constrain the half opening angle within 
the range of $N_{\rm H}^{\rm eq} = 10^{23}-10^{24}$ cm$^{-2}$,
the reflection strength $R_{\rm torus} = 0.45^{+0.13}_{-0.10}$ implies
$\theta_{\rm oa} \simeq 50^\circ-70^\circ$ by assuming $R_{\rm torus} \simeq {\rm cos(\theta_{\rm oa})}$. 
For NGC 4941, the half opening angle is estimated to be 
$\theta_{\rm oa} \simeq 60^\circ-70^\circ$ from 
the iron-K$\alpha$ equivalent width, which is 
consistent with $R_{\rm torus} = 0.64^{+0.69}_{-0.27}$.
Thus, in both targets, modest covering factors of the tori are
suggested.

According to the unified scheme, the torus covering fraction determines
the fraction of obscured AGNs. Thus, it is quite interesting to compare
with the result on the type-2 fraction derived from unbiased AGN
surveys. On the basis of the {\it Swift}/BAT survey performed in the
15--55 keV band, \citet{Bur11} suggest that the fraction has a peak of
$\approx$0.6--0.8 around $L_{\rm 15-55} = 10^{42-43}$ ergs s$^{-1}$ and
decreases toward both higher and lower luminosity ranges. At $L_{15-55}
= 10^{41-42}$ ergs s$^{-1}$, the fraction is estimated to be
$0.1-0.6$ (see e.g., their Figure~16), although the error is still quite
large due to a limited number of LLAGNs detected in the hard X-ray
survey. Similar results are also obtained from X-ray follow-up
observations of {\it IRAS} 12~$\mu$m selected galaxies \citep{Bri11b}. 
Based on the best-fit model, the 15--55 keV luminosity of NGC 1566 and 
NGC 4941 are $3.3\times10^{41}$ ergs s$^{-1}$ and $4.5\times10^{41}$ 
ergs s$^{-1}$. 
Our result on the torus half-opening angle of these sources, 
$\theta_{\rm oa} \simeq 50^\circ-70^\circ$, can be
converted into an obscured fraction of 0.34--0.64, 
which is fully consistent with the above survey results at a similar
luminosity range.

Near infrared reverberation mapping observations of nearby AGNs clearly show that the
inner radius of AGN tori scales as $L_{\rm bol}^{-0.5}$
\citep[e.g.,][]{Sug06}, which can be physically explained by dust
sublimation by AGN irradiation. This means that tori in LLAGNs have a
much smaller scale-height $h_{\rm s}$ at the innermost radius $r_{\rm
in}$ compared with more luminous AGNs, in order to explain the observed
small covering fraction (i.e., small $h_{\rm s}/r_{\rm in}$). The
physical reason is not clear, in particular if it is ultimately related
to the low mass-accretion rates or just a consequence of radiation
effects. For a hydrostatic disk where the central gravity by the SMBH
dominates over its self-gravity, the ratio between ``thermal'' energy
(including kinetic energy of turbulent motion) and gravitational
potential determines the height to radius ratio.
For simplicity, let us assume that LLAGNs in the local universe have
similar black hole masses as normal AGNs but their Eddington ratios
are smaller. In fact, our objects have ${\rm log} L_{\rm bol}/L_{\rm
Edd} \approx -2.5$ (see Section~1), which are small compared with a
majority of {\it Swift}/BAT AGNs \citep{Win09}.
Then, for the same thermal energy inside the torus, the
ratio $h_{\rm s}/r_{\rm in}$ becomes smaller in LLAGNs, which could
explain the observed trend. Further studies of a larger sample of LLAGNs
covering a wide range of luminosity and Eddington ratio would be useful
to confirm our results and to understand the origin.

\acknowledgments
We acknowledge the usage of the HyperLeda data base (http://leda.univ-lyon1.fr).
Part of this work was financially supported by the Grant-in-Aid for JSPS
Fellows for young researchers (FT) and for Scientific Research 23540265
(YU) and 21244017 (YT), and by the grant-in-aid for the Global COE Program
``The Next Generation of Physics, Spun from Universality and Emergence''
from the Ministry of Education, Culture, Sports, Science and Technology
(MEXT) of Japan.

%%%%%%%%%%%%%%%%%%%%%%%%%

\end{document}